\documentclass[letterpaper]{article} 
\usepackage[submission]{aaai25}  
\usepackage{times}  
\usepackage{helvet}  
\usepackage{courier}  
\usepackage[hyphens]{url}  
\usepackage{graphicx} 
\usepackage{xcolor}
\usepackage{array} 
\usepackage{amsmath}
\usepackage{amsfonts} 

\usepackage{bm} 
\usepackage{natbib}  
\usepackage{caption} 
\usepackage{algorithm}
\usepackage{algorithmic}

\usepackage{newfloat}
\usepackage{listings}
\DeclareCaptionStyle{ruled}{labelfont=normalfont,labelsep=colon,strut=off} 
\floatstyle{ruled}
\newfloat{listing}{tb}{lst}{}
\floatname{listing}{Listing}

\title{Convolutional Deep Operator Networks for Learning Nonlinear Focused Ultrasound Wave Propagation in Heterogeneous Spinal Cord Anatomy}
\author{
    Avisha Kumar\textsuperscript{\rm 1},
    Xuzhe Zhi\textsuperscript{\rm 1},
    Zan Ahmad\textsuperscript{\rm 1},
    Minglang Yin\textsuperscript{\rm 2},
    Amir Manbachi\textsuperscript{\rm 1,2}
}
\affiliations{
    \textsuperscript{\rm 1}Johns Hopkins University, Baltimore, MD, USA\\
    \textsuperscript{\rm 2} Johns Hopkins University School of Medicine, Baltimore, MD, USA\\    
}

\begin{document}

\maketitle

\begin{abstract}
Focused ultrasound (FUS) therapy is a promising tool for optimally targeted treatment of spinal cord injuries (SCI), offering submillimeter precision to enhance blood flow at injury sites while minimizing impact on surrounding tissues. However, its efficacy is highly sensitive to the placement of the ultrasound source, as the spinal cord's complex geometry and acoustic heterogeneity distort and attenuate the FUS signal. Current approaches rely on computer simulations to solve the governing wave propagation equations and compute patient-specific pressure maps using ultrasound images of the spinal cord anatomy. While accurate, these high-fidelity simulations are computationally intensive, taking up to hours to complete parameter sweeps, which is impractical for real-time surgical decision-making. To address this bottleneck, we propose a convolutional deep operator network (DeepONet) to rapidly predict FUS pressure fields in patient spinal cords. Unlike conventional neural networks, DeepONets are well equipped to approximate the solution operator of the parametric partial differential equations (PDEs) that govern the behavior of FUS waves with varying initial and boundary conditions (i.e., new transducer locations or spinal cord geometries) without requiring extensive simulations. Trained on simulated pressure maps across diverse patient anatomies, this surrogate model achieves real-time predictions with only a 2\% loss on the test set, significantly accelerating the modeling of nonlinear physical systems in heterogeneous domains. By facilitating rapid parameter sweeps in surgical settings, this work provides a crucial step toward precise and individualized solutions in neurosurgical treatments.
\end{abstract}

\begin{links}
    \link{Code and Dataset}{https://github.com/avishakumar21/nonlinear-fus-with-neural-operators}
\end{links}

\section{Introduction}
Spinal cord injury (SCI) often leads to significant physiological issues following the initial trauma, including prolonged reduced blood flow to the injury site (hematoma) \cite{quadri2020recent}. Current treatments, like surgical decompression, cannot precisely restore blood flow to the hematoma for optimal healing \cite{lonjaret2014optimal, ahuja2017traumatic}. Focused ultrasound (FUS) therapy offers a targeted solution by generating acoustic pressure to enhance localized blood flow directly to a focal region \cite{hwang2021ultrasound, tsehay2022advances, morishita2014effects, hong2022ultrasound}. When administering FUS therapy, clinicians must ensure that the focal region of the resulting FUS pressure distribution maximally overlaps with the hematoma with minimal exposure to surrounding tissue for optimal therapeutic benefit. However, predicting FUS pressure fields in the spinal cord relative to transducer positions is difficult due to its complex geometric variations and acoustic heterogeneity that cause beam distortion \cite{kumar2023computational}. 
\par
\subsection{Computational Bottleneck}
Computer simulations that numerically solve the governing FUS wave propagation equations can determine the pressure distribution through patient-specific spinal cord anatomy using ultrasound images \cite{treeby2018rapid, kumar2023patient}. This pressure distribution is crucial for non-invasively inferring the biological effects of a specific transducer placement, especially when treating a sensitive organ such as the spinal cord. To achieve optimal therapeutic results, all possible source locations along the region of interest in the spine must be evaluated. However, these predictions cannot be made prior to surgery, as a laminectomy is required to remove surrounding vertebral bone and create an acoustic window for ultrasound imaging. Other imaging methods, such as magnetic resonance imaging (MRI) and computed tomography (CT), do not adequately capture all the soft tissues within the spinal cord that have varying acoustic properties, which are essential for accurate simulations. Since these simulations can take several minutes to hours per patient, it is important to develop a faster method for solving the FUS wave equations in patient-specific spinal cord to support timely intraoperative decision-making.
\par
Previous works in predicting ultrasound behavior includes physics-informed neural networks (PINNs) on ultrasound acoustic wave data to simulate regions with surface-breaking cracks and PINNs for transcranial ultrasound wave propagation \cite{wang2023physics, shukla2020physics}. These methods embed physical constraints of the acoustic wave equation into the loss equation of their models to ensure that predictions align closely with the underlying physics, improving accuracy in scenarios where direct measurements or high-quality imaging data may be limited \cite{raissi2019physics}. While these approaches address the limitations of data-driven models that require extensive, diverse training datasets and overcome the challenges of traditional mesh-based methods—such as time-intensive calculations for high-dimensional problems—they must be retrained for each new input domain. This retraining is impractical for FUS treatment planning, as each patient’s spinal cord geometry is unique, especially after sustaining an injury. 
\par
\subsection{Operator Learning for PDEs} 
Neural operator learning is a framework for learning mappings between infinite-dimensional function spaces, such as the solution generators for systems of partial differential equations (PDEs). This approach differs from traditional neural networks, which are designed to learn mappings between finite-dimensional vector spaces. By learning the underlying relationships between input functions (initial and boundary conditions) and their corresponding PDE solutions, neural operators aim to generalize across families of PDEs without requiring retraining for each new configuration \cite{kovachki2023neural}.

Mathematically, a neural operator \(\mathcal{G}_{\theta}\) maps an input function space \(\mathcal{U}(\Omega_{\alpha}; \mathbb{R}^{d_u})\) to an output function space \(\mathcal{V}(\Omega_{\alpha}; \mathbb{R}^{d_v})\), as follows:
\begin{equation}
\label{NO}
\mathcal{G}_{\theta}: \mathcal{U}(\Omega_{\alpha}; \mathbb{R}^{d_u}) \to \mathcal{V}(\Omega_{\alpha}; \mathbb{R}^{d_v}),
\end{equation}
where \(\Omega_{\alpha} \subset \mathbb{R}^d\) represents the spatial domain and \(\alpha \in \mathcal{A}\) parametrizes its shape. The parameters of the neural operator, \(\theta \in \Theta\), are optimized during training. The dimensions \(d_u\) and \(d_v\) refer to the sizes of the input and output function spaces, which may be subspaces of Sobolev spaces or spaces of continuous functions. These function spaces provide a broad mathematical setting for applications where the PDE solutions vary smoothly or have certain regularity properties.

The true solution operator \(\mathcal{G}\) is approximated by \(\mathcal{G}_{\theta}\) using training data in the form of input-output pairs, \(\{u_i, v_i\}_{i=1}^N\), where \(u_i \in \mathcal{U}\) represents an input function, and \(v_i = \mathcal{G}(u_i) \in \mathcal{V}\) is the corresponding PDE solution. These pairs are typically generated from high-fidelity numerical simulations of the governing equations. For example, in modeling focused ultrasound (FUS) pressure fields in the spinal cord, \(\mathcal{U}\) may represent the initial and boundary conditions of the acoustic wave equation in a patient-specific spinal cord geometry, while \(\mathcal{V}\) corresponds to the resulting pressure distribution in the tissue.

By learning a direct mapping from inputs to solutions, neural operators can predict the solution to a PDE without solving it numerically, offering significant computational advantages. This makes them well-suited for applications requiring rapid predictions, such as real-time surgical planning for FUS treatment.

\subsection{Proposed Framework}
The proposed network architecture used in this work combines convolutional neural networks (CNNs) \cite{anwar2018medical} and deep operator networks (DeepONets) \cite{lu2021learning} to learn a generalizable mapping between images of the spinal cord anatomy and transducer/source location (input functions) and the resulting pressure fields across the spatial domain after FUS sonication (output functions). Trained on simulated pressure maps from diverse patient anatomies, the proposed model predicts pressure distributions across varying spinal cord geometries in real-time, achieving just 2\% error on the test set. While operator learning has been explored in medical contexts \cite{loeffler2024graph, zhou2024ai, yin2022simulating}, this paper introduces its application for real-time optimization of neurosurgical interventions. 

\section{Methods}

\subsection{Data Generation}
Learning a surrogate model to accelerate simulations requires an abundantly diverse and expressive training dataset, with both input images and corresponding ground truth simulation results. To train our neural operator, we generated simulated pressure maps using 1,000 sagittal B-mode ultrasound images of the thoracic spinal cord from 25 porcine subjects, captured both before and after contusion injuries \cite{kumar2024novel}. Each image ($25.6 mm \times 8.1 mm$) displayed anatomical structures with varying acoustic properties, including the dorsal space, dura, cerebrospinal fluid (CSF), pia, spinal cord, ventral space, and injury site (Figure \ref{dataset}A). Ground truth masks for these soft-tissue regions were used to define the computational domain to generate patient-specific acoustic phantoms of the spinal cords (Figure \ref{dataset}B). For some images, due to the acquisition angle or the severity of injury, the anatomical boundaries between the dura, CSF, and pia or between the ventral dura and ventral space were indistinct. In these cases, the regions were grouped and labeled as the dura/pia complex or dura/ventral complex, respectively.

\begin{table*}[htb]
\centering
\caption{Acoustic properties of spinal cord anatomies of interest \cite{hasgall2022itis}}
\label{tab:acoustic properties}
\begin{tabular}{|>{\centering\arraybackslash}m{3.5cm}|>{\centering\arraybackslash}m{2cm}|%
>{\centering\arraybackslash}m{2cm}|>{\centering\arraybackslash}m{2.1cm}|>{\centering\arraybackslash}m{2cm}|}
\hline
\textbf{Material} & \textbf{Sound Speed (m/s)} & \textbf{Density (kg/m$^3$)} & \textbf{Attenuation Constant (dB/MHz$^y$cm)} & \textbf{Acoustic Nonlinearity (B/A)} \\ \hline
Dorsal Space         & 1578.2 & 1050  & 0.2152  & 6.11 \\ \hline
Dura                 & 1500.0 & 1174  & 1.1641  & 6.72 \\ \hline
CSF                  & 1504.5 & 1007  & 0.0087  & 4.96 \\ \hline
Pia                  & 1500.0 & 1174  & 1.1641  & 6.72 \\ \hline
Dura/Pia Complex     & 1500.0 & 1174  & 1.1641  & 6.72 \\ \hline
Spinal Cord          & 1542.0 & 1075  & 0.778   & 6.72 \\ \hline
Hematoma             & 1560.1 & 1062.5 & 0.4968 & 6.42 \\ \hline
Dura/Ventral Complex & 2064.0 & 1339  & 1.77642 & 7.00 \\ \hline
Ventral Space        & 2577.2 & 1504  & 2.38875 & 7.28 \\ \hline
\end{tabular}
\end{table*}

\subsection{Computer Simulations}
We use \texttt{k-wave}, a MATLAB toolbox for modeling acoustic wave propagation, to generate the simulated pressure maps \cite{treeby2018rapid}. By solving a system of first-order partial differential equations (PDEs) computationally equivalent to the generalized Westervelt equation, \texttt{k-Wave} is designed to account for nonlinearities introduced to wave propagation caused by high magnitude acoustic waves common in biomedical ultrasonics. The framework also models acoustically heterogeneous media, such as the spinal cord, where sound speed and ambient density vary spatially. The governing PDEs (Equations \ref{eq:momentum_conservation} -- \ref{eq:pressure_density_relation}) are solved at each time step via pseudo-spectral methods to obtain the pressure of the system at each grid location, 
\begin{align}
\frac{\partial \mathbf{u}}{\partial t} &= -\frac{1}{\rho_0} \nabla p, \label{eq:momentum_conservation} \\
\frac{\partial \rho}{\partial t} &= - (2\rho + \rho_0) \nabla \cdot \mathbf{u} - \mathbf{u} \cdot \nabla \rho_0, \label{eq:mass_conservation} \\
p &= c_0^2 \left( \rho + \bm{d} \cdot \nabla \rho_0 + \frac{B}{2A} \frac{\rho^2}{\rho_0} - L \rho \right), \label{eq:pressure_density_relation}
\end{align}
where $\mathbf{u}$ is the acoustic particle velocity, $\rho$ is the acoustic density, $p$ is the acoustic pressure, $\rho_0$ is ambient density, $c_0$ is the sound speed, and $d$ is the acoustic particle displacement. $L$ is a linear integro-differential operator used in \texttt{k-Wave} that accounts for acoustic absorption and dispersion.

The acoustic properties for each anatomical region were obtained with a literature search for physiological accuracy and are summarized in Table \ref{tab:acoustic properties} \cite{hasgall2022itis}. Due to limited data in the literature, the following assumptions were made. The hematoma properties were estimated by averaging the acoustic values of the spinal cord and blood to represent the effects of ruptured blood vessels during injury. The pia is assumed to have the same acoustic properties as the dura because of their similar echogenicity. Acoustic nonlinearity parameters for the dura, CSF, and spinal cord were derived from analogous biological tissues. Specifically, the dura, pia, and spinal cord were given the nonlinearity parameter of brain tissue, while the CSF was modeled using the nonlinearity parameter of water. The ventral space itself was assumed to exhibit properties averaging those of bone and cartilage, and the dura/ventral complex was assumed to be average values of dura and ventral space. The attenuation constant for all these media was defined by a power-law absorption exponent (y) of 1.05, indicating that dispersion effects were insignificant in the simulations.
 \par 
After calibrating our patient-specific computational grids (mesh size: $512 \times 162$ grid points, element spacing: $5e-5$) to exhibit acoustic properties of the spinal cord, a single element transducer emitting a continuous sine wave was configured based on characteristics of typical FUS transducers. The operating frequency of the transducer was set to 2.5 MHz, with a sonication time of 1e-4 seconds and a focal length of 5 mm. The source location was systematically varied across 8 positions, spaced 3 mm apart along the sagittal axis of the spine. The corresponding pressure distributions were recorded at each position, providing a comprehensive discretization of possible pressure maps across the entire imaged anatomy.

\begin{figure}[htb]
\centerline{\includegraphics[scale = 0.55]{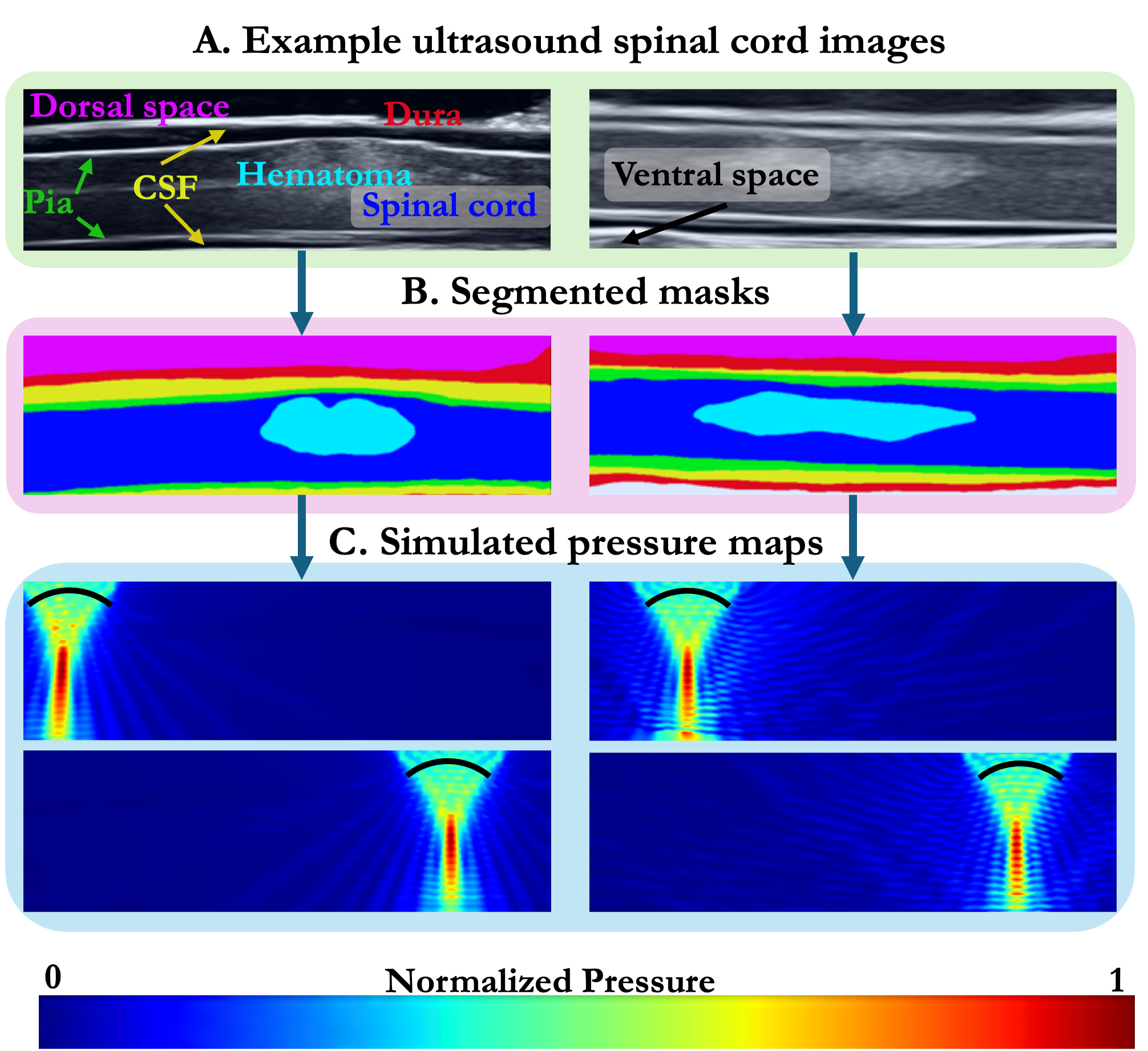}}
\caption{Example ultrasound spinal cord images in the dataset along with their segmented masks and visualizations of corresponding pressure maps at various source locations.} 
\label{dataset}
\end{figure}

\par 
These pressure maps (Figure \ref{dataset}C) serve as the ground truth simulation dataset for the proposed neural operator model, which was divided into training, validation, and test sets with an 80-10-10 split. There was no overlap of patient spinal cord images between the sets to prevent data leakage and overfitting. 
\subsection{Network Architecture}
Our proposed network architecture features a convolutional DeepONet, which is well suited for our problem given that we have a fixed rectangular domain $\Omega$ for all training samples. In this context, we allow the pressure distributions from the FUS simulations to depend on 2 critical parameters: 
\begin{enumerate}
    \item The unique spinal cord geometry of each patient which prescribes the acoustic heterogeneity in the parametric PDE.
    \item The specific transducer locations at which the pressure field is evaluated.
\end{enumerate}

\begin{figure*}[htb]
\centerline{\includegraphics[scale = 0.45]{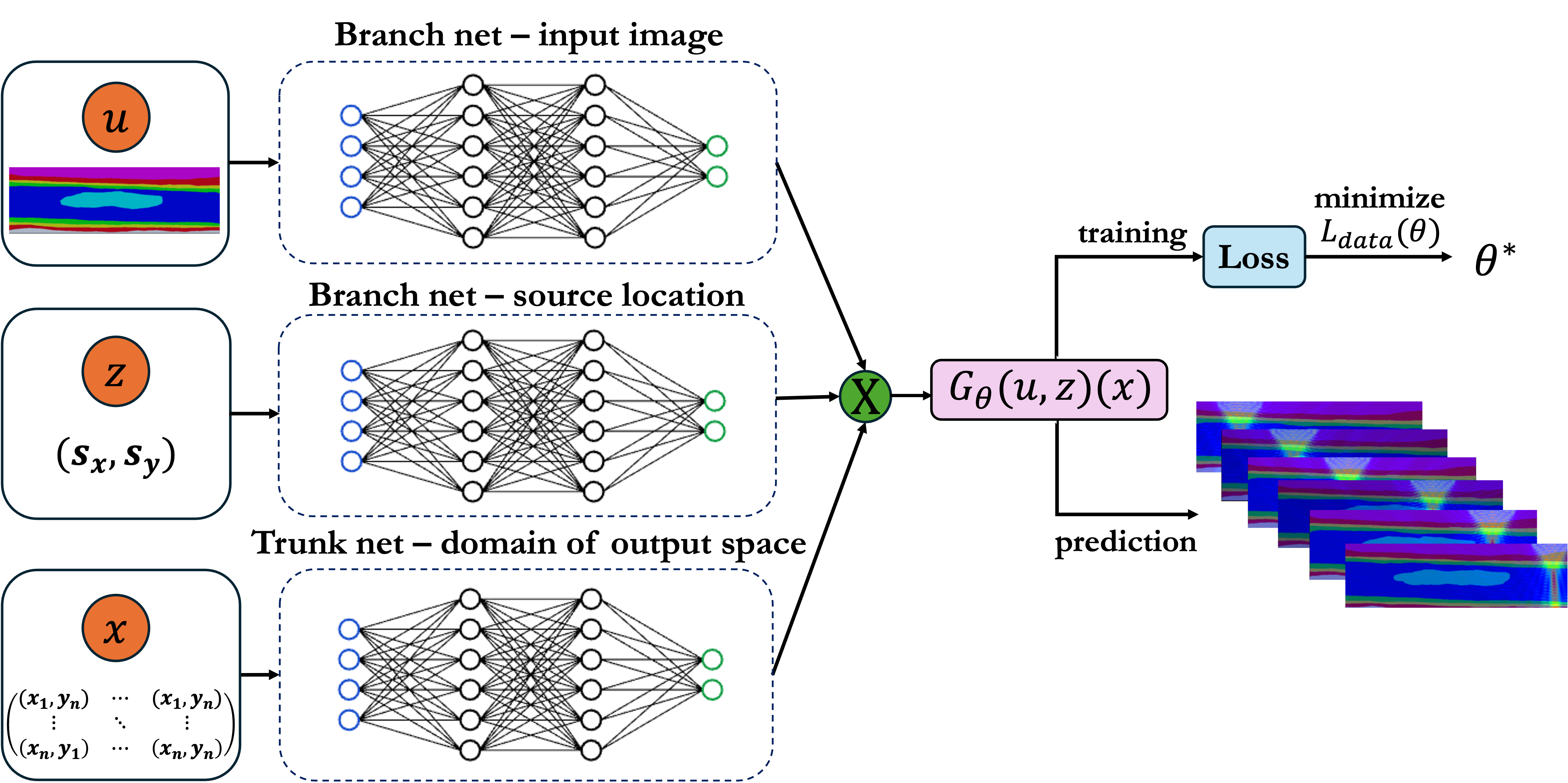}}
\caption{Visualization of proposed model architecture. This model consists of two branch nets for encoding the input function space (spinal cord image and transducer location) and a trunk net for encoding the domain of the output functions (discrete coordinates at which they are evaluated). The outputs of these deep neural networks are then merged with a dot product to approximate the true solution generator \(\mathcal{G}\) with the neural operator $\mathcal{G}_{\theta}$. The loss is minimized to obtain the optimal set of parameters ($\theta$) for the solution operator during the training process.} 
\label{deeponet}
\end{figure*}

To account for these dependencies, our model incorporates two branch networks to encode the input function space and one trunk network to encode the output domain. The branch networks process patient-specific geometric and source-specific information, while the trunk network encodes the grid locations of the output pressure distribution.

Let the governing wave equation be represented succinctly by:
\[
\mathcal{L}(p; \bm{u},  \mathbf{z}) = f, \quad \text{in } \Omega,
\]
where:
\begin{itemize}
\item \(p \in \mathcal{V}(\Omega)\) is the acoustic pressure distribution,
\item \(\bm{u} \in \mathcal{U}_1\) represents the patient-specific spinal cord geometry (e.g., anatomical masks),
\item \(\mathbf{z} \in \mathcal{U}_2\) represents the transducer location parameters (e.g., \((x, y)\)),
\item \(\Omega \subset \mathbb{R}^2\) is the rectangular computational domain.
\end{itemize}
The neural operator approximates the solution operator \(\mathcal{G}: \mathcal{U}_1 \times \mathcal{U}_2 \to \mathcal{V}\), such that:
\[
p = \mathcal{G}(\bm{u},\mathbf{z}).
\]

The complete network architecture (depicted in Figure \ref{deeponet}) consists of the following components:
\begin{itemize}
    \item \textbf{Branch Network 1:} A CNN with three convolutional layers, which encodes the patient-specific spinal cord geometry \(\bm{u}\). The CNN processes segmented anatomical masks from ultrasound images and outputs a latent representation \(\mathbf{g}_{\text{geo}} \in \mathbb{R}^{q}\).
    \item \textbf{Branch Network 2:} A fully connected neural network (FCNN) with three hidden layers, which encodes the transducer location \(\mathbf{z}\). This network outputs a latent representation \(\mathbf{g}_{\text{src}} \in \mathbb{R}^{r}\).
    \item \textbf{Trunk Network:} A fully connected neural network (FCNN) with three hidden layers, which encodes the grid locations \(\mathbf{x} \in \Omega\). The output of the trunk network is a spatial representation \(\mathbf{h}_{\text{spatial}} \in \mathbb{R}^{s}\).
\end{itemize}

The outputs of the branch and trunk networks are combined using a Hadamard product, resulting in a continuous representation of the predicted pressure map:
\[
p(\mathbf{x}) \approx \mathcal{G}_{\theta}(\bm{u}, \mathbf{z})(\mathbf{x}) = \sum_{i=1}^n (\mathbf{g}_{\text{src}, i} \cdot \mathbf{g}_{\text{geo}, i}) \cdot \mathbf{h}_{\text{spatial}, i}(\mathbf{x}).
\]

The model is trained using simulation data generated by \texttt{k-Wave}, with the goal of minimizing the relative \(L_2\) loss:
\[
\mathcal{L} = \frac{\|p_{\text{pred}} - p_{\text{true}}\|_2}{\|p_{\text{true}}\|_2}.
\]

\begin{table*}[htb]
\caption{Characteristics of deep learning models for approximating focused ultrasound pressure distribution in patient-specific spinal cord anatomy.}
\label{model-description}
\centering
\begin{tabular}{|c|c|c|c|c|c|c|}
\hline
\textbf{Model}                                                      & \textbf{\begin{tabular}[c]{@{}c@{}}Trainable\\ Parameters\end{tabular}} & \textbf{\begin{tabular}[c]{@{}c@{}}Learning \\ Rate\end{tabular}} & \textbf{\begin{tabular}[c]{@{}c@{}}Batch \\ Size\end{tabular}} & \textbf{\begin{tabular}[c]{@{}c@{}}Training \\ Epochs\end{tabular}} & \textbf{Optimizer} & \textbf{\begin{tabular}[c]{@{}c@{}}Loss \\ function\end{tabular}} \\ \hline
\begin{tabular}[c]{@{}c@{}}Proposed Neural\\ Operator\end{tabular}                                         & 5,286,656                                                                  & 0.001                                                             & 4                                                              & 1000                                                                & Adam               & Relative $L_2$                                                       \\ \hline
Baseline CNN                                                        & 23,649                                                                  & 0.01                                                              & 32                                                             & 100                                                                 & Adam               & Relative $L_2$                                                       \\ \hline
Baseline FCN  & 22,833,473                                                                  & 0.01                                                              & 32                                                             & 100                                                                 & Adam               & Relative $L_2$                                                       \\ \hline
\end{tabular}
\end{table*}

The Adam optimizer with a step-based learning rate schedule is employed for training. By efficiently approximating the operator \(\mathcal{G}\), this architecture maps patient-specific anatomical features and transducer parameters to acoustic pressure distributions. This enables rapid parameter sweeps, facilitating optimal source placement for treatment.

\begin{figure*}[htb]
\centerline{\includegraphics[scale = 0.55]{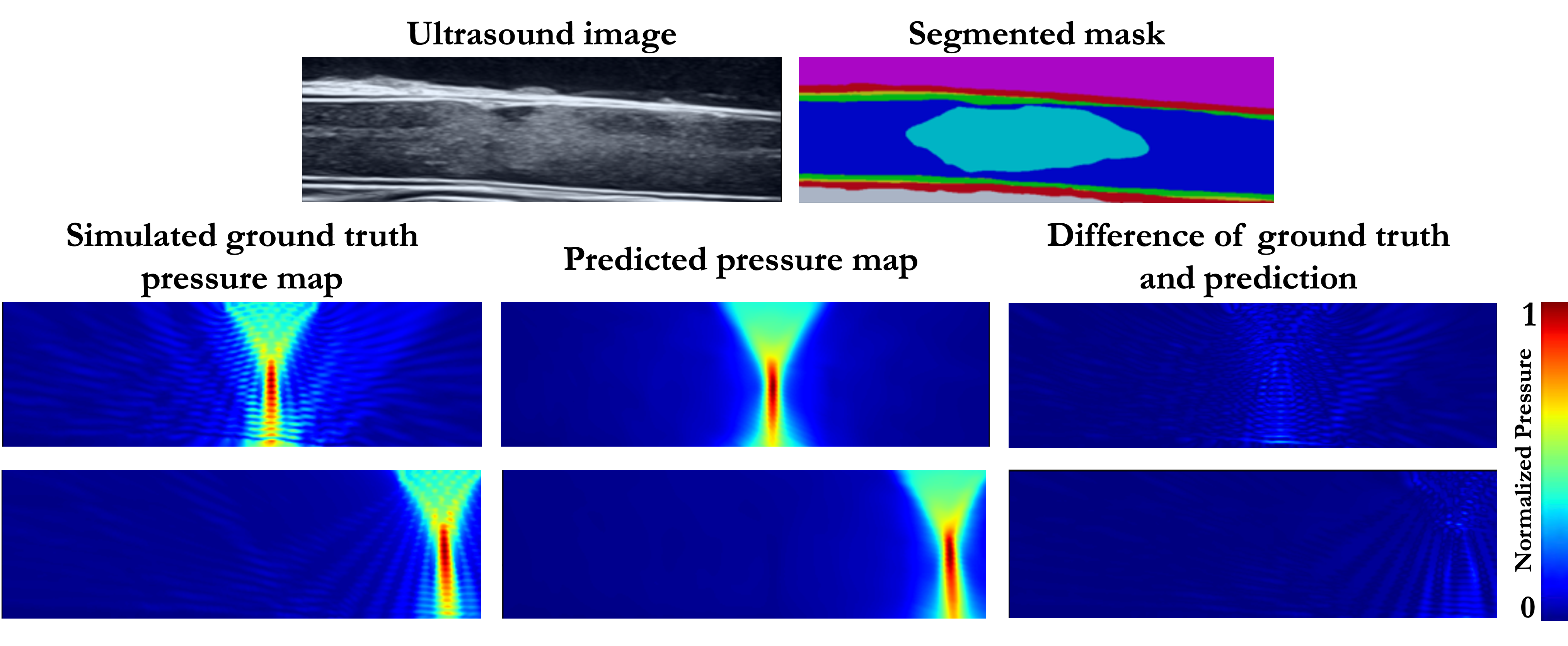}}
\caption{Visualization of results from the proposed neural operator model, including the ultrasound image of the injured spinal cord, the corresponding segmented mask used as input for the neural operator, high-fidelity ground truth pressure maps at source locations 4 and 8 obtained from \texttt{k-Wave}, the model’s predicted pressure maps for those locations, and the difference between the predictions and the ground truth.} 
\label{result_fig}
\end{figure*}

We compare the proposed operator network against two baseline models to evaluate performance in predicting pressure maps in heterogeneous spinal cord anatomy. The first baseline is a CNN model, similar to the branch net in the DeepONet architecture, consisting of three convolutional layers with additional layers for upsampling the output to match the dimensions of the expected simulation. The second baseline is a Fully Convolutional Network (FCN) with regression, a deep learning architecture typically designed for dense, pixel-wise prediction tasks, such as semantic segmentation, where continuous values are predicted for each pixel \cite{long2015fully, sofka2017fully}. Our FCN is a pretrained ResNet34 with upsampling layers for dimension matching. These models were trained on a Windows 11 Machine (8 GB RAM) with 24 GB NVIDIA GeForce RTX 3090 graphics unit and 14th Gen Intel Core i5-14600 processor (14 cores, 20 threads, 2.7GHZ to 5.2GHz turbo frequency). The details of these models are provided in Table \ref{model-description}, and the hyperparameters are tuned to optimize performance for each model.

\section{Results}
\begin{figure}
\centerline{\includegraphics[scale = 0.24]{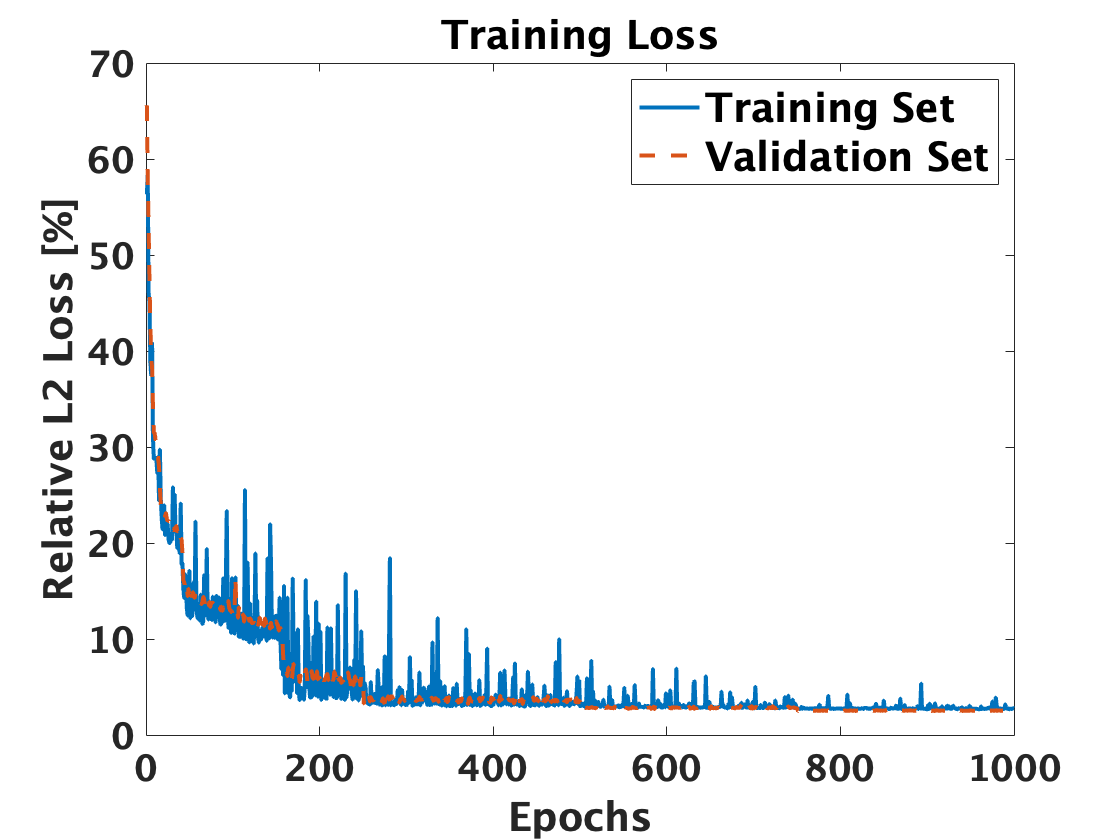}}
\caption{The loss for the training and validation set recorded at each training epoch for the proposed neural operator model.} 
\label{loss}
\end{figure}
\begin{figure}
\centerline{\includegraphics[scale = 0.49]{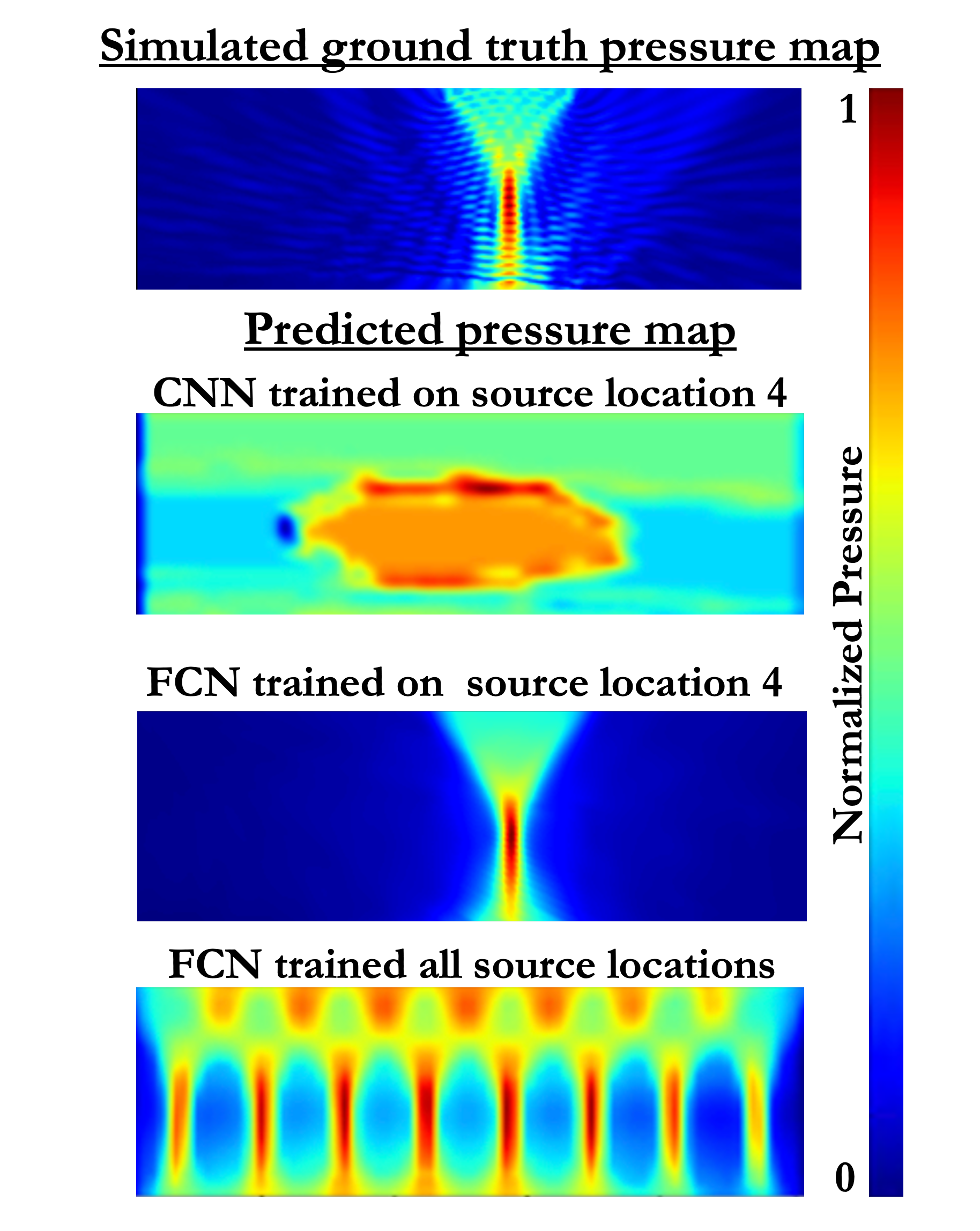}}
\caption{Visualization of the predicted pressure maps from the baseline comparison models for source location 4, including results from a convolutional neural network (CNN) trained only on source location 4, and fully convolutional networks (FCNs) trained on source location 4 and all source locations.} 
\label{benchmark}
\end{figure}

After developing the simulation training data and the task-specific deep learning models for predicting FUS pressure maps, each model was evaluated on unseen patient spinal cord geometries in the test set. The results are summarized in Table \ref{result_table}, with our proposed neural operator model outperforming the other deep learning models with only a 2\% loss in the test set evaluated against all 8 locations (Figure \ref{loss}). The visualization of the predictions made by the proposed neural operator model is presented in Figure \ref{result_fig}. We also assessed the performance of CNN and FCN models trained on a single source location as a baseline comparison against the DeepONet model. From the test loss and prediction results (Figure \ref{benchmark}), it is evident that the baseline CNN model fails to learn FUS pressure distributions in new spinal cord anatomy, with a loss of 64\% even when trained on a single source location. Although the baseline FCN model trained on a single source location achieves a low loss of 2\%, it proves impractical for location sweeps without embedding the source location into its architecture (69\% loss). 
\par

\begin{table}
\centering
\caption{Performance comparison of models trained on either all source locations or on a single source location (position 4) for generating pressure maps in heterogeneous patient-specific spinal cords.}
\label{result_table}
\begin{tabular}{|c|c|}
\hline
\textbf{Model and Task}                                                                   & \textbf{\begin{tabular}[c]{@{}c@{}}Relative L2 Loss\\ on Validation Set\end{tabular}} \\ \hline
\begin{tabular}[c]{@{}c@{}}Proposed neural operator - \\ All source location\end{tabular} & \textbf{0.024}                                                                        \\ \hline
\begin{tabular}[c]{@{}c@{}}Baseline CNN - \\ Source location 4\end{tabular}               & 0.647                                                                                 \\ \hline
\begin{tabular}[c]{@{}c@{}}Baseline FCN - \\ Source location 4\end{tabular}               & 0.024                                                                                 \\ \hline
\begin{tabular}[c]{@{}c@{}}Baseline FCN - \\ All source locations\end{tabular}            & 0.687                                                                                 \\ \hline
\end{tabular}
\end{table}

\begin{table}
\centering
\caption{Time comparison between methods for generating 8 pressure maps across a heterogeneous patient-specific spinal cord.}
\label{time}
\begin{tabular}{|c|c|}
\hline
\textbf{Method}                                                      & \textbf{\begin{tabular}[c]{@{}c@{}}Time for generating 8 \\ pressure maps\end{tabular}} \\ \hline
\begin{tabular}[c]{@{}c@{}} Proposed neural \\ operator \end{tabular}                                             & 0.05 seconds                                                                                    \\ \hline
\begin{tabular}[c]{@{}c@{}}\texttt{k-Wave} \\ (numerical solver)\end{tabular} & 76.1 minutes                                                                                    \\ \hline
\end{tabular}
\end{table}

Additionally, we compare the time required to generate 8 pressure maps using predictions from the proposed model versus high-fidelity numerical solutions computed with \texttt{k-Wave} on a macOS system equipped with an Apple M1 chip, featuring an 8-core CPU with 4 performance cores and 4 efficiency cores, 8-core GPU, and 8 GB of unified memory. This benchmarking provides insights into the computational efficiency of the model on consumer-grade hardware. Our findings indicate that the inference on our model takes 0.05 seconds, while obtaining the pressure maps using the numerical solver \texttt{k-Wave}, takes 76.1 total minutes (Table \ref{time}).

\section{Discussion}
Numerical modeling is a promising tool for approximating the behavior of physical systems in complex media by solving the governing physical equations iteratively in a specified computational grid. These simulations allow clinicians to infer the biological effects of FUS in patient-specific anatomy, providing valuable insight in individualized treatment planning (i.e., where to place the source). However, the computational demands of these simulations limit their utility for real-time parameter sweeps needed to optimize intraoperative decision-making. High-fidelity ultrasound simulations using traditional finite difference and finite element methods require a resolution of 10 mesh points per acoustic wavelength \cite{zingg1996high}. This results in prohibitively large computational grids, especially when modeling high-frequency biomedical transducers operating in the MHz range. This challenge is particularly significant in the context of spinal cord injury, where preoperative ultrasound imaging is not feasible until vertebral bone removal, requiring simulations to be performed intraoperatively. While MRI and CT imaging are available, they fail to provide adequate soft-tissue delineation required for the acoustic solvers. Ultrasound imaging, on the other hand, inherently captures the acoustic differences within the spinal cord, which are crucial for accurately predicting the propagation of the nonlinear FUS beam.
\par
Our proposed neural operator architecture, which integrates CNNs and DeepONets, demonstrates the ability to learn the solution operator for nonlinear FUS wave propagation in heterogeneous, patient-specific spinal cord anatomy, enabling real-time prediction of pressure maps. Unlike traditional neural networks which excel at finding patterns in data, operator learning frameworks provide a more principled approach that results in a model which respects the underlying  continuous structure of the true PDE solution generator. This allows for predictions to be made based on various input functions without need for retraining or running expensive simulations repeatedly. In this case, the nonlinear operator is the mapping from a space of functions (patient spinal cord geometry) to another space of functions (acoustic pressure distribution). Accelerating FUS simulations can be transformative across several medical domains, such as glioblastoma or deep vein thrombosis, offering a novel approach to optimize ultrasound transducer placements for targeting tumors or blood clots while minimizing exposure to adjacent healthy tissue \cite{zhou2011high}. 
\par 
While a FCN can learn pressure distributions across the spinal cord for a single source location with model loss comparable to the proposed approach, it lacks the ability to generalize effectively across multiple source locations. As a result, it is unsuitable for parameter sweeps to determine the optimal source location. To achieve comparable generalization capabilities with this architecture, it would require training with source location embeddings, likely increasing the computational and resource burden. This limitation is further exacerbated by the fact that the baseline FCN model is more than 4 times larger than the proposed neural operator (Table \ref{model-description}). 
\par
Our proposed model efficiently and accurately learns the relationship between patient-specific anatomy and the corresponding \textit{in vivo} pressure distribution for various transducer placements. The online computing time of this network is over \textbf{\textit{91,000}} times faster than traditional simulations (Table \ref{time}), demonstrating the practical applicability of operator learning in surgical medicine, particularly for optimizing the use of therapeutic ultrasound. 
\par
Given the similarities in soft-tissue morphology, vasculature, and immune response after injury between porcine and human spinal cord anatomy, we anticipate that this model can generalize effectively to the human spinal cord \cite{toossi2021comparative}. Fine-tuning the final layers with human data would further enhance its performance and applicability. Moving forward, we aim to enhance this model by integrating a segmentation model into the preprocessing pipeline, replacing the use of the soft-tissue masks as inputs. This improvement will streamline intraoperative decision-making by eliminating the need for the tedious and manual segmentation currently required. Future efforts will focus on adapting the model to accept raw ultrasound images as inputs to the branch network instead of segmented masks, while maintaining high performance. Additionally, we plan to expand the training and evaluation process to include human spinal cord images, enhancing the clinical utility of this tool.

\section{Conclusion}
The deployment of advanced computational models in predictive ultrasound modeling holds potential to revolutionize personalized treatment planning by dramatically enhancing time efficiency and enabling scalable parameter sweeps at a fraction of the computational cost. Traditional numerical models, while highly accurate and physics-based, face a critical limitation: each new input, such as transducer location or patient anatomy, requires a complete recomputation of the solution, a computational luxury that cannot be afforded in time-sensitive environments like the operating room. In this paper, we introduced a deep operator network capable of predicting nonlinear focused ultrasound pressure maps in patient-specific, heterogeneous spinal cord anatomy, offering a paradigm-shifting approach to accelerating intraoperative treatment planning. With the advent of high-resolution ultrasound imaging and the increasing interest in focused ultrasound therapy for noninvasive treatments, the application of deep learning to develop efficient surrogate models can have far-reaching implications in healthcare, including tumor ablation and blood clot removal. This work underscores the immense promise of AI-driven solutions to address critical challenges in clinical decision-making and treatment optimization.

\section{Acknowledgments}
Amir Manbachi acknowledges funding support from the Defense Advanced Research Projects Agency (DARPA) (award \#N660012024075), along with the Johns Hopkins Institute for Clinical and Translational Research (ICTR)’s Clinical Research Scholars Program (KL2), administered by the National Center for Advancing Translational Sciences (NCATS), National Institute of Health (NIH), as well as the National Science Foundation (NSF)’s Small Businesses Technology Transfer (STTR) (award \#1938939). 

\bibliography{aaai25}

\begin{thebibliography}{26}
\providecommand{\natexlab}[1]{#1}

\bibitem[{Ahuja et~al.(2017)Ahuja, Wilson, Nori, Kotter, Druschel, Curt, and Fehlings}]{ahuja2017traumatic}
Ahuja, C.~S.; Wilson, J.~R.; Nori, S.; Kotter, M.; Druschel, C.; Curt, A.; and Fehlings, M.~G. 2017.
\newblock Traumatic spinal cord injury.
\newblock \emph{Nature reviews Disease primers}, 3(1): 1--21.

\bibitem[{Anwar et~al.(2018)Anwar, Majid, Qayyum, Awais, Alnowami, and Khan}]{anwar2018medical}
Anwar, S.~M.; Majid, M.; Qayyum, A.; Awais, M.; Alnowami, M.; and Khan, M.~K. 2018.
\newblock Medical image analysis using convolutional neural networks: a review.
\newblock \emph{Journal of medical systems}, 42: 1--13.

\bibitem[{Hasgall et~al.(2022)Hasgall, Di~Gennaro, Baumgartner, Neufeld, Lloyd, Gosselin, Payne, Klingenböck, and Kuster}]{hasgall2022itis}
Hasgall, P.; Di~Gennaro, F.; Baumgartner, C.; Neufeld, E.; Lloyd, B.; Gosselin, M.; Payne, D.; Klingenböck, A.; and Kuster, N. 2022.
\newblock IT’IS Database for thermal and electromagnetic parameters of biological tissues.

\bibitem[{Hong et~al.(2022)Hong, Lee, Park, Han, Kim, and Park}]{hong2022ultrasound}
Hong, Y.-r.; Lee, E.-h.; Park, K.-s.; Han, M.; Kim, K.-T.; and Park, J. 2022.
\newblock Ultrasound stimulation improves inflammatory resolution, neuroprotection, and functional recovery after spinal cord injury.
\newblock \emph{Scientific Reports}, 12(1): 3636.

\bibitem[{Hwang et~al.(2021)Hwang, Mampre, Ahmed, Suk, Anderson, Manbachi, and Theodore}]{hwang2021ultrasound}
Hwang, B.~Y.; Mampre, D.; Ahmed, A.~K.; Suk, I.; Anderson, W.~S.; Manbachi, A.; and Theodore, N. 2021.
\newblock Ultrasound in traumatic spinal cord injury: a wide-open field.
\newblock \emph{Neurosurgery}, 89(3): 372--382.

\bibitem[{Kovachki et~al.(2023)Kovachki, Li, Liu, Azizzadenesheli, Bhattacharya, Stuart, and Anandkumar}]{kovachki2023neural}
Kovachki, N.; Li, Z.; Liu, B.; Azizzadenesheli, K.; Bhattacharya, K.; Stuart, A.; and Anandkumar, A. 2023.
\newblock Neural operator: Learning maps between function spaces with applications to pdes.
\newblock \emph{Journal of Machine Learning Research}, 24(89): 1--97.

\bibitem[{Kumar et~al.(2024)Kumar, Kotkar, Jiang, Bhimreddy, Davidar, Weber-Levine, Krishnan, Kerensky, Liang, Leadingham et~al.}]{kumar2024novel}
Kumar, A.; Kotkar, K.; Jiang, K.; Bhimreddy, M.; Davidar, D.; Weber-Levine, C.; Krishnan, S.; Kerensky, M.~J.; Liang, R.; Leadingham, K.~K.; et~al. 2024.
\newblock A novel open-source ultrasound dataset with deep learning benchmarks for spinal cord injury localization and anatomical segmentation.
\newblock \emph{arXiv preprint arXiv:2409.16441}.

\bibitem[{Kumar et~al.(2023{\natexlab{a}})Kumar, Punnoose, Leadingham, Kerensky, Theodore, Thakor, and Manbachi}]{kumar2023patient}
Kumar, A.; Punnoose, J.; Leadingham, K. M.~K.; Kerensky, M.~J.; Theodore, N.; Thakor, N.~V.; and Manbachi, A. 2023{\natexlab{a}}.
\newblock A Patient-specific Preplanning Treatment Algorithm for Focused Ultrasound Therapy of Spinal Cord Injury.
\newblock In \emph{2023 11th International IEEE/EMBS Conference on Neural Engineering (NER)}, 1--4. IEEE.

\bibitem[{Kumar et~al.(2023{\natexlab{b}})Kumar, Tsehay, Gonzalez, Kerensky, Bell, Theodore, Thakor, and Manbachi}]{kumar2023computational}
Kumar, A.; Tsehay, Y.; Gonzalez, E.; Kerensky, M.~J.; Bell, M. A.~L.; Theodore, N.; Thakor, N.~V.; and Manbachi, A. 2023{\natexlab{b}}.
\newblock Computational modeling towards focused ultrasound therapy for spinal cord injury: visualization of beam propagation through patient-specific anatomy.
\newblock In \emph{Medical Imaging 2023: Image-Guided Procedures, Robotic Interventions, and Modeling}, volume 12466, 276--282. SPIE.

\bibitem[{Loeffler et~al.(2024)Loeffler, Ahmad, Ali, Yamamoto, Popescu, Yee, Lal, Trayanova, and Maggioni}]{loeffler2024graph}
Loeffler, S.~E.; Ahmad, Z.; Ali, S.~Y.; Yamamoto, C.; Popescu, D.~M.; Yee, A.; Lal, Y.; Trayanova, N.; and Maggioni, M. 2024.
\newblock Graph Fourier Neural Kernels (G-FuNK): Learning Solutions of Nonlinear Diffusive Parametric PDEs on Multiple Domains.
\newblock \emph{arXiv preprint arXiv:2410.04655}.

\bibitem[{Long, Shelhamer, and Darrell(2015)}]{long2015fully}
Long, J.; Shelhamer, E.; and Darrell, T. 2015.
\newblock Fully convolutional networks for semantic segmentation.
\newblock In \emph{Proceedings of the IEEE conference on computer vision and pattern recognition}, 3431--3440.

\bibitem[{Lonjaret et~al.(2014)Lonjaret, Lairez, Minville, and Geeraerts}]{lonjaret2014optimal}
Lonjaret, L.; Lairez, O.; Minville, V.; and Geeraerts, T. 2014.
\newblock Optimal perioperative management of arterial blood pressure.
\newblock \emph{Integrated blood pressure control}, 49--59.

\bibitem[{Lu et~al.(2021)Lu, Jin, Pang, Zhang, and Karniadakis}]{lu2021learning}
Lu, L.; Jin, P.; Pang, G.; Zhang, Z.; and Karniadakis, G.~E. 2021.
\newblock Learning nonlinear operators via DeepONet based on the universal approximation theorem of operators.
\newblock \emph{Nature machine intelligence}, 3(3): 218--229.

\bibitem[{Morishita et~al.(2014)Morishita, Karasuno, Yokoi, Morozumi, Ogihara, Ito, Fujiwara, Fujimoto, and Abe}]{morishita2014effects}
Morishita, K.; Karasuno, H.; Yokoi, Y.; Morozumi, K.; Ogihara, H.; Ito, T.; Fujiwara, T.; Fujimoto, T.; and Abe, K. 2014.
\newblock Effects of therapeutic ultrasound on intramuscular blood circulation and oxygen dynamics.
\newblock \emph{Journal of the Japanese Physical Therapy Association}, 17(1): 1--7.

\bibitem[{Quadri et~al.(2020)Quadri, Farooqui, Ikram, Zafar, Khan, Suriya, Claus, Fiani, Rahman, Ramachandran et~al.}]{quadri2020recent}
Quadri, S.~A.; Farooqui, M.; Ikram, A.; Zafar, A.; Khan, M.~A.; Suriya, S.~S.; Claus, C.~F.; Fiani, B.; Rahman, M.; Ramachandran, A.; et~al. 2020.
\newblock Recent update on basic mechanisms of spinal cord injury.
\newblock \emph{Neurosurgical Review}, 43: 425--441.

\bibitem[{Raissi, Perdikaris, and Karniadakis(2019)}]{raissi2019physics}
Raissi, M.; Perdikaris, P.; and Karniadakis, G.~E. 2019.
\newblock Physics-informed neural networks: A deep learning framework for solving forward and inverse problems involving nonlinear partial differential equations.
\newblock \emph{Journal of Computational physics}, 378: 686--707.

\bibitem[{Shukla et~al.(2020)Shukla, Di~Leoni, Blackshire, Sparkman, and Karniadakis}]{shukla2020physics}
Shukla, K.; Di~Leoni, P.~C.; Blackshire, J.; Sparkman, D.; and Karniadakis, G.~E. 2020.
\newblock Physics-informed neural network for ultrasound nondestructive quantification of surface breaking cracks.
\newblock \emph{Journal of Nondestructive Evaluation}, 39: 1--20.

\bibitem[{Sofka et~al.(2017)Sofka, Milletari, Jia, and Rothberg}]{sofka2017fully}
Sofka, M.; Milletari, F.; Jia, J.; and Rothberg, A. 2017.
\newblock Fully convolutional regression network for accurate detection of measurement points.
\newblock In \emph{Deep Learning in Medical Image Analysis and Multimodal Learning for Clinical Decision Support: Third International Workshop, DLMIA 2017, and 7th International Workshop, ML-CDS 2017, Held in Conjunction with MICCAI 2017, Qu{\'e}bec City, QC, Canada, September 14, Proceedings 3}, 258--266. Springer.

\bibitem[{Toossi et~al.(2021)Toossi, Bergin, Marefatallah, Parhizi, Tyreman, Everaert, Rezaei, Seres, Gatenby, Perlmutter et~al.}]{toossi2021comparative}
Toossi, A.; Bergin, B.; Marefatallah, M.; Parhizi, B.; Tyreman, N.; Everaert, D.~G.; Rezaei, S.; Seres, P.; Gatenby, J.~C.; Perlmutter, S.~I.; et~al. 2021.
\newblock Comparative neuroanatomy of the lumbosacral spinal cord of the rat, cat, pig, monkey, and human.
\newblock \emph{Scientific Reports}, 11(1): 1955.

\bibitem[{Treeby et~al.(2018)Treeby, Budisky, Wise, Jaros, and Cox}]{treeby2018rapid}
Treeby, B.~E.; Budisky, J.; Wise, E.~S.; Jaros, J.; and Cox, B. 2018.
\newblock Rapid calculation of acoustic fields from arbitrary continuous-wave sources.
\newblock \emph{The Journal of the Acoustical Society of America}, 143(1): 529--537.

\bibitem[{Tsehay et~al.(2022)Tsehay, Weber-Levine, Kim, Chara, Alomari, Awosika, Liu, Ehresman, Lehner, Hwang et~al.}]{tsehay2022advances}
Tsehay, Y.; Weber-Levine, C.; Kim, T.; Chara, A.; Alomari, S.; Awosika, T.; Liu, A.; Ehresman, J.; Lehner, K.; Hwang, B.; et~al. 2022.
\newblock Advances in monitoring for acute spinal cord injury: a narrative review of current literature.
\newblock \emph{The Spine Journal}, 22(8): 1372--1387.

\bibitem[{Wang et~al.(2023)Wang, Wang, Liang, Li, Zeng, and Liu}]{wang2023physics}
Wang, L.; Wang, H.; Liang, L.; Li, J.; Zeng, Z.; and Liu, Y. 2023.
\newblock Physics-informed neural networks for transcranial ultrasound wave propagation.
\newblock \emph{Ultrasonics}, 132: 107026.

\bibitem[{Yin et~al.(2022)Yin, Ban, Rego, Zhang, Cavinato, Humphrey, and Em~Karniadakis}]{yin2022simulating}
Yin, M.; Ban, E.; Rego, B.~V.; Zhang, E.; Cavinato, C.; Humphrey, J.~D.; and Em~Karniadakis, G. 2022.
\newblock Simulating progressive intramural damage leading to aortic dissection using DeepONet: an operator--regression neural network.
\newblock \emph{Journal of the Royal Society Interface}, 19(187): 20210670.

\bibitem[{Zhou et~al.(2024)Zhou, Wan, Huang, Li, Peng, Anandkumar, Brady, Sternberg, and Daraio}]{zhou2024ai}
Zhou, T.; Wan, X.; Huang, D.~Z.; Li, Z.; Peng, Z.; Anandkumar, A.; Brady, J.~F.; Sternberg, P.~W.; and Daraio, C. 2024.
\newblock AI-aided geometric design of anti-infection catheters.
\newblock \emph{Science Advances}, 10(1): eadj1741.

\bibitem[{Zhou(2011)}]{zhou2011high}
Zhou, Y.-F. 2011.
\newblock High intensity focused ultrasound in clinical tumor ablation.
\newblock \emph{World journal of clinical oncology}, 2(1): 8.

\bibitem[{Zingg, Lomax, and Jurgens(1996)}]{zingg1996high}
Zingg, D.~W.; Lomax, H.; and Jurgens, H. 1996.
\newblock High-accuracy finite-difference schemes for linear wave propagation.
\newblock \emph{SIAM Journal on Scientific Computing}, 17(2): 328--346.

\end{thebibliography}

\end{document}